# THE CONTROL SYSTEM OF THE AEgIS EXPERIMENT AT CERN

## System sterowania eksperymentu AEgIS w CERN

Modern atomic and quantum physics experiments require complex sequences and interaction with many components to run the measurements. In addition, the timing should be synchronised to the duration of the observed processes, often in the order of a ns or even below. The experimental setups are generally not permanent and are subject to operation conditions and configuration changes. These features must be reflected in a flexible and customisable control system. The steady increase in complexity has made custom solutions progressively challenging to maintain. Also, such methods are likely to have bottlenecks when implementing complex sequences, which can be challenging under changing conditions of the experiment. To overcome such problems, the AEgIS experiment (Antimatter Experiment, gravity, interferometry, and spectroscopy) at CERN adopted Sinara and ARTIQ as the backbone of the autonomous control system. This decision settled the long-term support and scalability and fulfilled all the requirements of a frontier science experiment such as AEgIS.

The main experimental goal of AEgIS is to measure the ballistic fall of a cold antihydrogen horizontal pulsed beam [1–3]. The antihydrogen is formed using the charge exchange reaction between Rydberg excited positronium (Ps) and cold antiprotons previously loaded and cooled down using sympathetic cooling. Once the reaction happens, the Rydberg-excited antihydrogen escapes the confinement in the Penning-Malmberg trap. It is ready for further manipulations for spectroscopic measurements or determination of its coupling to gravity.

Ph.D., D.Sc. Georgy Kornakov, M.Sc. Jakub Zieliński,
Ph.D. Grzegorz Kasprowicz,
Warsaw University of Technology, Warsaw, Poland

**ABSTRACT**

The AEgIS experiment at CERN recently decided to adopt a control system solution based on the Sinara/ARTIQ open hardware and software infrastructure. This decision meant to depart from the previously used paradigm of custom-made electronics and software to control the experiment's equipment. Instead, adopting a solution with long-term support and used in many quantum physics experiments guarantees a vivid community using similar infrastructures. This transition reduces the risks and development timeline for integrating new equipment seamlessly within the setup. This work reviews the motivation, the setup, and the chosen hardware and presents several planned further steps in developing the control system.

**KEYWORDS:** Control System, Sinara, ARTIQ, antimatter, ion trap

**STRESZCZENIE**

Eksperyment AEgIS w CERN niedawno podjął decyzję o przyjęciu rozwiązania systemu kontroli opartego na infrastrukturze otwartego sprzętu i oprogramowania Sinara/ARTIQ. Ta decyzja oznacza odejście od dotychczasowego paradygmatu wykorzystywania spersonalizowanej elektroniki i oprogramowania do sterowania sprzętem eksperymentu. Zamiast tego, przyjęcie rozwiązania zapewniającego długoterminowe wsparcie i stosowanego w szeregu eksperymentów z fizyki kwantowej gwarantuje aktywną społeczność korzystającą z podobnych infrastruktur. Ta przemiana zmniejsza ryzyko oraz skraca terminy rozwoju w celu bezproblemowej integracji nowego sprzętu w konfiguracji. Niniejsza praca analizuje motywację, konfigurację, wybrane elementy sprzętu oraz prezentuje kilka planowanych kolejnych kroków w rozwoju systemu kontroli.

**SŁOWA KLUCZOWE:** systemy sterowania, Sinara, ARTIQ, antymateria, pułapka jonowa

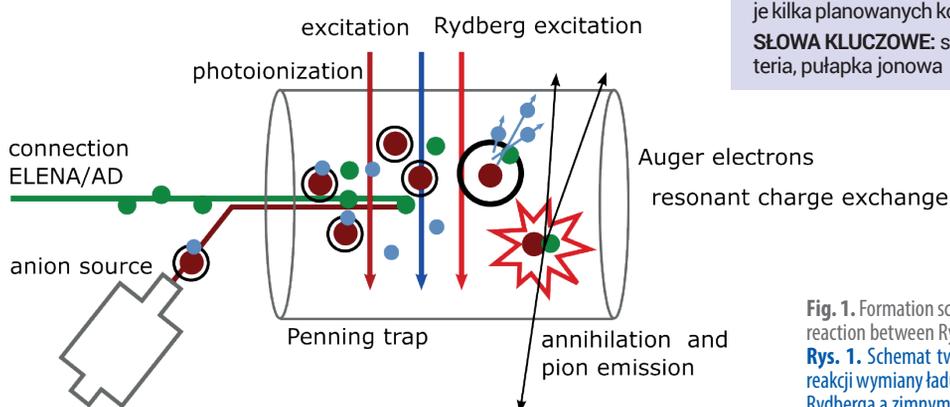

**Fig. 1.** Formation scheme of antiprotonic atoms using the charge-exchange reaction between Rydberg excited trapped atoms and cold antiprotons
**Rys. 1.** Schemat tworzenia atomów antyprotonowych z wykorzystaniem reakcji wymiany ładunku między uwięzionymi atomami wzbudzonymi przez Rydberga a zimnymi antyprotonami

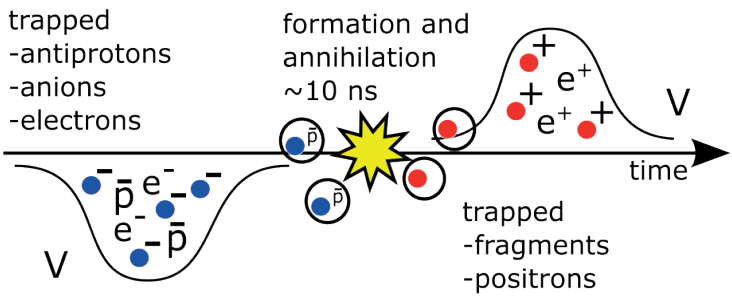

**Fig. 2.** Scheme for re-capturing and cooling the highly charged ions produced after the annihilation of antiprotons on the surface of the capturing nucleus.
**Rys. 2.** Schemat ponownego wychwytywania i schładzania silnie naładowanych jonów powstałych po anihilacji antyprotonów na powierzchni jądra przechwytującego

The same trapping system can also produce other types of exotic atoms in a pulsed form containing bound matter and antimatter [4]. Although antimatter and matter annihilate when they are in close contact, they can create, for a short time, about $\mu$s, metastable atomic systems where, for example, an electron is substituted by an antiproton. These systems are the subject of study as they can test the existing theories using accurate spectroscopic techniques. Figure 1 shows the scheme for producing pulsed antiprotonic atoms as a starting point for the process of trapped negative ions and cold antiprotons. Once they are all mixed, a three-level laser system brings the negative ions to the Rydberg state first by the photodetachment (IR laser) of the excess electron and then in a two-step process from the ground state (UV) toward highly excited Rydberg state (Visible frequency). Then, a spontaneous charge-exchange reaction happens, and the 2000 heavier antiproton substitutes the Rydberg electron. After cascading down, it reaches the nucleus's surface, and annihilation occurs. In general, only a few nucleons will be evaporated, the remaining heavy fragment with low kinetic energy. In Figure 2, it is shown the process of re-capturing the remainder fragment by reversing the polarity of the trap. The subsequent sympathetic cooling can be achieved using positron plasma, radiating excess heat through synchrotron radiation.

An advanced control system to do these experiments is required to orchestrate lasers, HV electrodes, pulsers, acquisition systems, triggers from the accelerator, and other devices such as actuators and cameras. In the following, we introduce the Sinara system and the control for the electrodes of the Penning-Malmberg trap of the AEgIS experiment.

## CONTROL SYSTEM

Sinara is an open-source hardware ecosystem [5] thoughtfully designed by physicists and explicitly tailored to quantum science laboratories and experiments, particularly those engaged in research with trapped ions. This platform presents a modular and rigorously tested set of FPGA-based solutions tailored to atomic, molecular, and optical (AMO) experiments with an exceptional sub-nanosecond time resolution. Its core aspiration lies in accommodating the distinctive requirements of individual laboratories while constantly bracing the assumptions of top architectural design, reproducibility, and comprehensive support to the final users who do not need to be experts in FPGA programming or electronics to perform operations with the help of the control system based on Sinara hardware.

Sinara employs two principal hardware form factors: the microTCA (uTCA) and the Eurocard Extension Modules (EEM). Figure 3 illustrates a Sinara setup used as a test bench for the AEgIS traps.

Functioning in tandem with Sinara, the ARTIQ control software (Advanced Real-Time Infrastructure for Quantum Physics) [6] constitutes a pivotal component of this ecosystem. ARTIQ is a high-level, Python-based programming language permitting researchers to describe the entire experimental environment using familiar code structures. This description is subsequently compiled and executed on FPGA hardware, granting deterministic nanosecond-level timing precision and sub-microsecond latency. A user-friendly interface and code clarity complement this formidable performance, thus ensuring an optimal user experience and long-term maintenance of the projects.

By unifying the strengths of the Sinara and ARTIQ systems, researchers can comprehensively manage and oversee the entirety of their data acquisition and electronics framework.

> **ARTIQ IS A HIGH-LEVEL, PYTHON-BASED PROGRAMMING LANGUAGE PERMITTING RESEARCHERS TO DESCRIBE THE ENTIRE EXPERIMENTAL ENVIRONMENT USING FAMILIAR CODE STRUCTURES.**

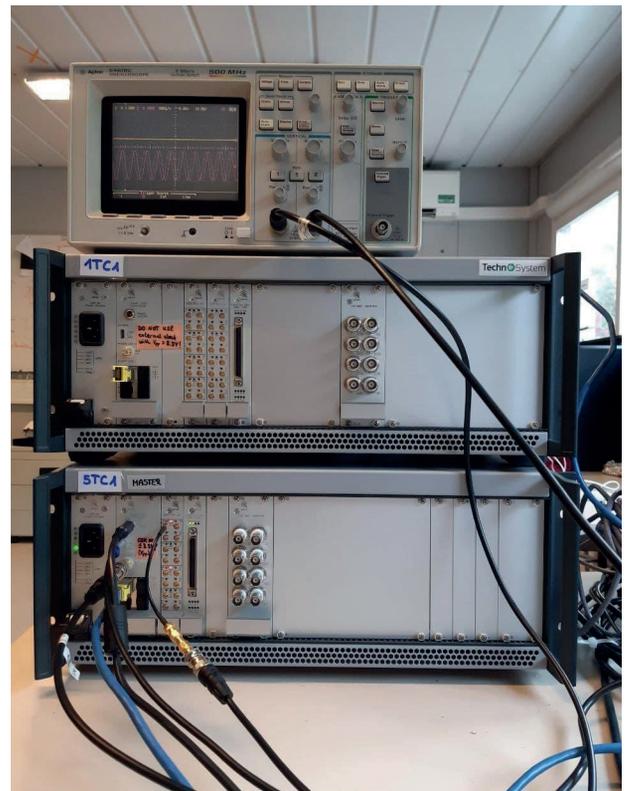

**Fig. 3.** Testing bench consisting of Sinara crates for the HV electrodes of the 5 and 1 T traps of AEgIS.
**Rys. 3.** Stanowisko testowe składające się ze skrzynek Sinara do elektrod WN pułapek 5 i 1 T firmy AEgIS.

## AEgIS TRAP CONTROL SYSTEM

The AEgIS trap control system has the following requirements [7]:
- Triggering DAC: level of ms
- Trigger response: level of μs
- Triggering synchronicity: level of ns

A new board to fulfil these characteristics has been incorporated into the Sinara family [8]. It is a High Voltage amplifier that can deliver an output range of +-200 V, as shown in *Figure 4*. The design has a dedicated 8-channel board which provides 1 MHz bandwidth and 50 R output impedance. Other features of the board are the capability of a quick output disconnection of the line controlled via EEM using OptoMos to limit the noise in the system and additional over-temperature protection for increased safety.

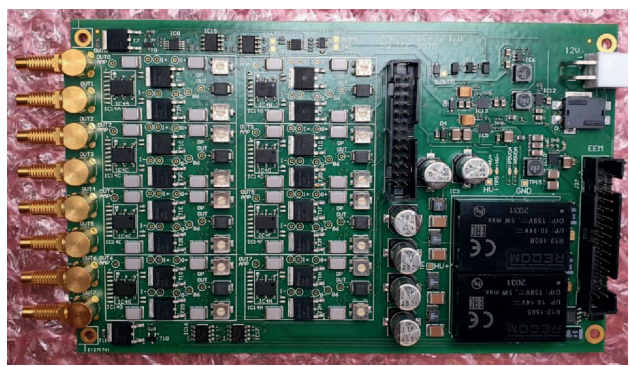

**Fig. 4.** HV amplifier with an 8-channel board designed to set the AEgIS traps' voltages. The board provides independent outputs from -200 to 200 V with 1 MHz bandwidth and 50 R output impedance. In addition, it has over-temp protection and quick output disconnect controlled via EEM using OptoMos to reduce the coupling of noise in the electrodes.
**Rys. 4.** Wzmacniacz wysokonapięciowy z płytą 8-kanałową specjalnie zaprojektowaną do ustawiania napięć na pułapkach AEgIS. Płytka zapewnia niezależne wyjścia od -200 do 200 V z szerokością pasma 1 MHz i impedancją wyjściową 50 R. Ponadto posiada zabezpieczenie przed przegrzaniem i szybkie odłączanie wyjścia kontrolowane przez EEM przy użyciu OptoMos w celu zmniejszenia sprzężenia szumów w elektrodach.

## FURTHER DEVELOPMENTS

Other subsystems of the AEgIS experiment are adopting the control system based on Sinara/ARTIQ. It will be the primary driver of the Paul trap-based negative ion source. This source is a key element to succeeding with the experiments to synthesise antiprotonic atoms. At the same time, it is planned to be used for electrostatic steering of the plasmas and charged particle beams in the apparatus. The system is constructed to allow online optimisation of the working parameters, maximising the loading efficiency and keeping charged particles inside the main traps.

## SUMMARY

The adoption of Sinara/ARTIQ by the AEgIS experiment at CERN has boosted the capability of the control system to perform complex AMO physics experiments which use antimatter. The reduced complexity of preparing the experimental sequences and the reproducibility of the results give a solid perspective for the experiment. The long-term support and the existence of a community using similar technical solutions strengthens the transmission of the accumulated know-how and allows us to invest more time in developing further experiments. ❖

### ACKNOWLEDGEMENTS

*This work is funded by the Research University – Excellence Initiative of Warsaw University of Technology via the strategic funds of the Priority Research Centre of High Energy Physics and Experimental Techniques and by the Polish National Science Centre under agreements no. 2022/45/B/ST2/02029, no. 2022/46/E/ST2/00255, and by the Polish Ministry of Education and Science under agreement no. 2022/WK/06.*